\documentclass[twocolumn,notitlepage,aps,prb,10pt,superscriptaddress,floatfix,letterpaper,reprint
]{revtex4-1}

\usepackage{hyperref}
\usepackage[usenames,dvipsnames]{color}
\usepackage{amsmath}
\usepackage[english]{babel}
\usepackage{amssymb}
\usepackage{graphicx}
\usepackage{epstopdf}
\usepackage[normalem]{ulem}
\usepackage{subfigure}
\usepackage{todonotes}
\usepackage{braket}

\usepackage[para]{threeparttable}

\definecolor{linkcolor}{HTML}{399B03}
\definecolor{urlcolor}{HTML}{399B03}
\hypersetup{pdfstartview=FitH, linkcolor=linkcolor,urlcolor=urlcolor,colorlinks=true}
\hypersetup{
    unicode=false,          
    pdftoolbar=true,        
    pdfmenubar=true,        
    pdffitwindow=false,     
    pdfstartview={FitH},    
    pdfauthor={},         
    colorlinks=true,        
    linkcolor=blue,         
    citecolor=red,          
}

\newcommand{\tk}{\mathbf{k}}
\newcommand{\tr}{\mathbf{r}}

\newcommand{\V}{\tilde{V}}

\newcommand{\Tr}[1]{\mathrm{Tr}\left[#1\right]}

\begin{document}

\title{Effect of propagator renormalization on the band gap of insulating solids}

\author{Sergei Iskakov}
\affiliation{%
 Department of Physics, University of Michigan, Ann Arbor, Michigan 48109, USA
}%
\author{Alexander A. Rusakov}%
\affiliation{%
 Department of Chemistry, University of Michigan, Ann Arbor, Michigan 48109, USA
}%
\author{Dominika Zgid}%
\affiliation{%
 Department of Chemistry, University of Michigan, Ann Arbor, Michigan 48109, USA
}%
\affiliation{%
 Department of Physics, University of Michigan, Ann Arbor, Michigan 48109, USA
}%
\affiliation{%
 Center for Computational Quantum Physics, Flatiron Institute, New York, NY 10010, USA
}%

\author{Emanuel Gull}%
\affiliation{%
 Department of Physics, University of Michigan, Ann Arbor, Michigan 48109, USA
}%
\affiliation{%
 Center for Computational Quantum Physics, Flatiron Institute, New York, NY 10010, USA
}%

\date{\today}

\begin{abstract}
We present momentum-resolved spectral functions and band gaps from bare and self-consistent second order perturbation theory for insulating periodic solids. We establish that, for systems with large gap sizes, both bare and self-consistent perturbation theory yield reasonable gaps. However, smaller gap sizes require a self-consistent adjustment of the propagator. In contrast to results obtained within a quasi-particle formalism used on top of bare second order perturbation theory, no unphysical behaviour of the band gap is observed. Our implementation of a fully self-consistent, $\Phi$-derivable and thermodynamically consistent finite temperature diagrammatic perturbation theory forms a framework on which embedding theories such as the dynamical mean field theory or self-energy embedding theories can be implemented.
\end{abstract}

\maketitle

\section{Introduction}
A truly ab-initio quantitative many-body description of weakly correlated systems beyond density functional theory (DFT)~\cite{RevModPhys.71.1253, Kohn65} remains challenging despite enormous theoretical \cite{Hedin65} and computational \cite{PhysRevB.34.5390,RevModPhys.74.601,doi:10.1021/ct500958p,DESLIPPE20121269,PhysRevB.74.035101,doi:10.1021/ct300835h} advances in the last thirty years.  While such calculations are important for explaining the behavior of insulating materials, they are also necessary as a first step for many strongly correlated computational techniques such as some derivatives~\cite{10.1007/1-4020-2708-7_4,PhysRevMaterials.1.043803, Zgid15,zgid_njp17,Tran_generalized_seet} of the dynamical mean field theory (DMFT)~\cite{Georges96,Kotliar06,doi:10.1080/00018730701619647}, where the first calculation step consists of evaluating the system by a perturbative weak coupling method. 

Three types of perturbation theories exist: bare (non-self-consistent) perturbation theory based on the expansion  of the original Hamiltonian in interaction terms, where neither propagators nor interactions are renormalized; self-consistent perturbation theories where propagators but not the interactions are renormalized; and perturbation theories with both propagators and interactions renormalized. 

Bare second order perturbation theory is known as M\o{}ller~-~Plesset second order (MP2)~\cite{Moller34,Wilberg86,Pople76,scuseria_mp2_solids,usvyat_MP2_general,usvyat_MP2_LiH,usvyat_MP2_molecular_crystals,hirata_fast_mp2_2009, Kresse_solids_mp2_I,Kresse_solidsII} when applied to real materials. 
The self-consistent second order Green's function perturbation theory (GF2) \cite{Phillips14,Phillips15,Rusakov16,Welden16,kananenka_hybrif_gf2,Dahlen05} renormalizes propagators but not interactions. While fully self-consistent GW~\cite{Hedin65} renormalizes both propagators and interactions, its non-self-consistent variants such as G$_0$W$_0$~\cite{PhysRevB.34.5390,AULBUR20001} (partially) renormalize interactions without renormalizing propagators.

The differences in the treatment of propagators and interactions between all these types of perturbation theories are crucial since approaches that do not renormalize the interactions are expected to fail in metallic 3D  systems~\cite{fetter2003quantum,Kresse_solidsII}. This breakdown is not expected to occur in insulators.

For weakly correlated materials, most results so far have been obtained within  approaches such as MP2~\cite{usvyat_MP2_LiH,usvyat_MP2_molecular_crystals,Kresse_solidsII,doi:10.1063/1.4961301} and GW \cite{PhysRevB.95.235123,KresseGW1,KresseGW2} at zero temperature. 
Only recently, finite-temperature results for solids have started to appear \cite{Kutepov12,Kutepov16,Kutepov17,Kutepov17b}. It was demonstrated that MP2 gaps are wildly inaccurate for materials with band gaps smaller than 
$6$ eV~\cite{Kresse_solidsII}, leading to a breakdown of the band gap estimation for silicon and silicon carbide. 
In contrast, zero-temperature, non-self-consistent GW~\cite{Hedin65} has been very successful in predicting band gaps for semiconductors. 
This success is usually attributed to the renormalization of the interactions by an infinite series of `bubble' (RPA \cite{Hedin65}) diagrams -- the same diagrams that render this method convergent in the metallic limit.

The renormalization of interactions and propagators is commonly discussed in many-body textbooks \cite{fetter2003quantum,Mattuck92}, usually at the example of weakly interacting or uniform systems. Due to both the computational cost and implementation difficulties, their effect in realistic solids is difficult to explore.
Since there is no obvious reason for the breakdown of the perturbative series in semiconductors and band insulators, it is interesting to explore how the renormalization with self-consistent propagators but unrenormalized interactions affects the band gaps and to compare these results to the ones obtained by MP2.
Moreover, since the MP2 band gap is evaluated using approximate band energies for the lowest unoccupied and highest occupied bands \cite{Kresse_solidsII} reminiscent of the formulas usually employed to solve the quasi-particle (QP) equations in GW~\cite{PhysRevB.76.165106}, it is interesting to compare these band gaps to gaps evaluated without solving the QP equations.

In this paper, we focus on quantifying the effects of the renormalization of propagators and self-energies on the value of band gaps in simple 3D solids. This is made possible by the implementation of a fully self-consistent finite temperature second order perturbation theory (GF2) that was so far not available for realistic 3D systems. Comparison between renormalized calculations, unrenormalized calculations, and experimental data also allows us to indirectly infer the contribution of higher order terms excluded from our calculations. Among these are the `RPA'-like screening terms included in the GW approximation.

\section{Method}\label{Sec:Method}
We investigate the physics of 3D solids by choosing a finite basis set on each atom, and choosing a finite lattice of atoms periodic in all three spatial directions. This yields a periodic electronic structure Hamiltonian best expressed in a Bloch basis in reciprocal space, resulting in the 
 Coulomb integrals 
\begin{align}
V_{ijkl}^{\tk_1\tk_2\tk_3\tk_4} = \iint  
\frac{\phi^{*}_{i\tk_1,\tr}\phi_{l\tk_4,\tr}\phi^{*}_{j\tk_2,\tr'}\phi_{k\tk_3,\tr'}}{|\tr-\tr'|} d\tr d\tr',
\label{Integral}
\end{align} 
where $\phi$ are basis functions in reciprocal space. Translational invariance is guaranteed by the momentum conservation $\tk_1+\tk_2 =
\tk_3+\tk_4$. These integrals can be decomposed \cite{doi:10.1063/1.1679012,doi:10.1002/qua.560120408,doi:10.1021/ct200352g,PhysRevB.71.073103}
into a product of two low-rank tensors,
$V_{ijkl}^{\tk_1\tk_2\tk_3\tk_4} = \sum_{Q}\tilde{V}_{i\tk_1,l\tk_4}^{Q}\tilde{V}_{j\tk_2,k\tk_3}^{Q}$, where $Q$ is an auxiliary basis index.
For the first order diagram we employ Ewald summation to treat the divergence at zero momentum. For the second-order diagram we excluded this point to
avoid the divergence of the exchange diagram.

The second-order self-energy is then evaluated in reciprocal space in an imaginary time formalism,
\begin{align}\label{eq:sigma2}
\Sigma^{\tk,(2)}_{ij}(\tau) &= -(2\V_{q\tk_1,j\tk}^{Q'} \V_{l\tk_2,n\tk_3}^{Q'} - \V_{l\tk_2,j\tk}^{Q'}\V_{q\tk_1,n\tk_3}^{Q'}) \\  \nonumber
\times \V_{i\tk,p\tk_1}^{Q}& \V_{m\tk_3,k\tk_2}^{Q}
G^{\tk_1}_{pq}(\tau)G^{\tk_2}_{kl}(\tau)G^{\tk_3}_{nm}(-\tau)\delta_{\tk+\tk_3,\tk_1+\tk_2},
\end{align}
with $\tau$ denoting imaginary time $0\leq\tau\leq\beta$ and $\beta=1/k_BT$ the inverse of the physical temperature $T$ (assuming Einstein summation over repeated indices).

In bare second-order perturbation theory, the Hartree-Fock Green's function is employed in Eq.~\ref{eq:sigma2} and the Dyson equation is evaluated only once to yield the interacting Green's function.
 
In renormalized perturbation theories such as GF2, to achieve self-consistency in both the density matrix (or, correspondingly, the  Fock matrix and the frequency independent term of the self-energy $\Sigma_\infty$) and the dynamical self-energy $\Sigma(\tau)$, we employ a modification of the iterative procedure described in Ref.~\onlinecite{Phillips14} for molecular systems.
First, we adjust the chemical potential to find the correct particle number of the Hartree-Fock (HF) solution. We then obtain the HF propagator, calculate the second-order self-energy, and recompute the interacting Green's function and density matrix using the Dyson equation, adjusting the chemical potential until the correct density is found. This propagator is then used for the next second-order self-energy evaluation, until convergence is achieved in all quantities.

Energies, entropies, free energies, and specific heats are then computed using standard thermodynamic formulae \cite{Luttinger60,PhysRevB.62.4858,fetter2003quantum,Dahlen06,Welden16}, for detailed derivations see appendix~\ref{appx:GP}.

Quantities obtained in diagrammatic approximations by thermodynamic integration are in general dependent on the integration path  \cite{Baym61,Baym62}; i.e. the integration of a quantity such as the energy or the entropy may differ if it is obtained by integration from zero $T$, infinite $T$, infinite chemical potential or via coupling constant integration. So-called `$\Phi$-derivable' \cite{Luttinger60}, self-consistent methods, such as the self-consistent GF2 method investigated here or the fully self-consistent GW approximation, avoid this problem and are intrinsically thermodynamically consistent.

Standard finite temperature perturbation theories are formulated on the imaginary axis, thus, static quantities such as the density matrix, the energy, the entropy, the static magnetic susceptibility or the specific heat are directly accessible. In contrast, real-frequency-dependent quantities such as the spectral function, the gap, the optical conductivity, or the dynamical magnetic susceptibility require analytical continuation to the real axis in order to be compared to experiment. 
This analytical continuation is ill conditioned and leads to an amplification of uncertainties, even for data known up to numerical precision, in particular at high temperature and high frequency. The problem is intrinsic to the finite-temperature field-theory formalism on the imaginary axis and can only be overcome by reformulating the method in frequency space or real time. In the present work, we used the maximum entropy method \cite{jarrell1996bayesian,LEVY2017149}. Other methods, such as the spectral method~\cite{Otsuki17}, Pad\'{e} \cite{Wang18}, stochastic analytical continuation~\cite{Mishchenko2000,pavarini2012correlated,Fuchs09}, or Consistent Constraints~\cite{Goulko17} could be explored, as could the continuation of the self-energy to obtain spectra and gaps via the quasi-particle equation~\cite{Wang09}.

\begingroup
\squeezetable
\begin{table}[bth]
\begin{threeparttable}[!ht]
\begin{tabular}{ | c || c | c | c |c | c  | c | c |}
System & HF\cite{Kresse_solidsII}& QPMP2\cite{Kresse_solidsII} & MP2 & GF2 & scGW\cite{KresseGW2} & QSGW\cite{Kutepov17} & exper. \\ 
 \hline
C      & 13.1 & 1.9 & 7.19 & 4.6   & 6.41 & 6.18  & 5.5\tnote{a} \\
LiH   & 11.2 & --  & 7.33 & 5.93  & --  & --   & 4.99\tnote{b}\\
MgO & 15.5 & 7.1 & 9.49 & 7.32  & 9.53 & 9.42  & 7.8\tnote{c}\\
LiF    & 21.8 & 14.2 & 13.03 & 13.03 & --  & 16.63 & 14.2\tnote{d}\\
Ne    & 25.3 & 20.3 & 20.55 & 20.55 & --  & --   & 21.7\tnote{f} \\
\end{tabular}
\begin{tablenotes}
\item[a] Refs.~\cite{CLARK1959481,PhysRev.136.A1445,PhysRev.170.683,Clark312,PhysRevLett.16.354}.
\item[b] Refs.~\cite{RID:0908160310813-98,RID:0908160310813-96}.
\item[c] Ref.~\cite{WHITED19731903}.
\item[d] Ref.~\cite{PhysRevB.13.5530}. 
\item[f] Refs.~\cite{Boursey70,PhysRevLett.34.528}.
\end{tablenotes}
\end{threeparttable}
\caption{\label{tab:cmp}Band gaps. Columns: numerical results as obtained in Refs.~\onlinecite{Kresse_solidsII,Kresse_solidsII,KresseGW2,Kutepov17} or as described in text. Last column: experiment.}
\end{table}
\endgroup

\begin{figure}[tbh]
\includegraphics[width=\columnwidth]{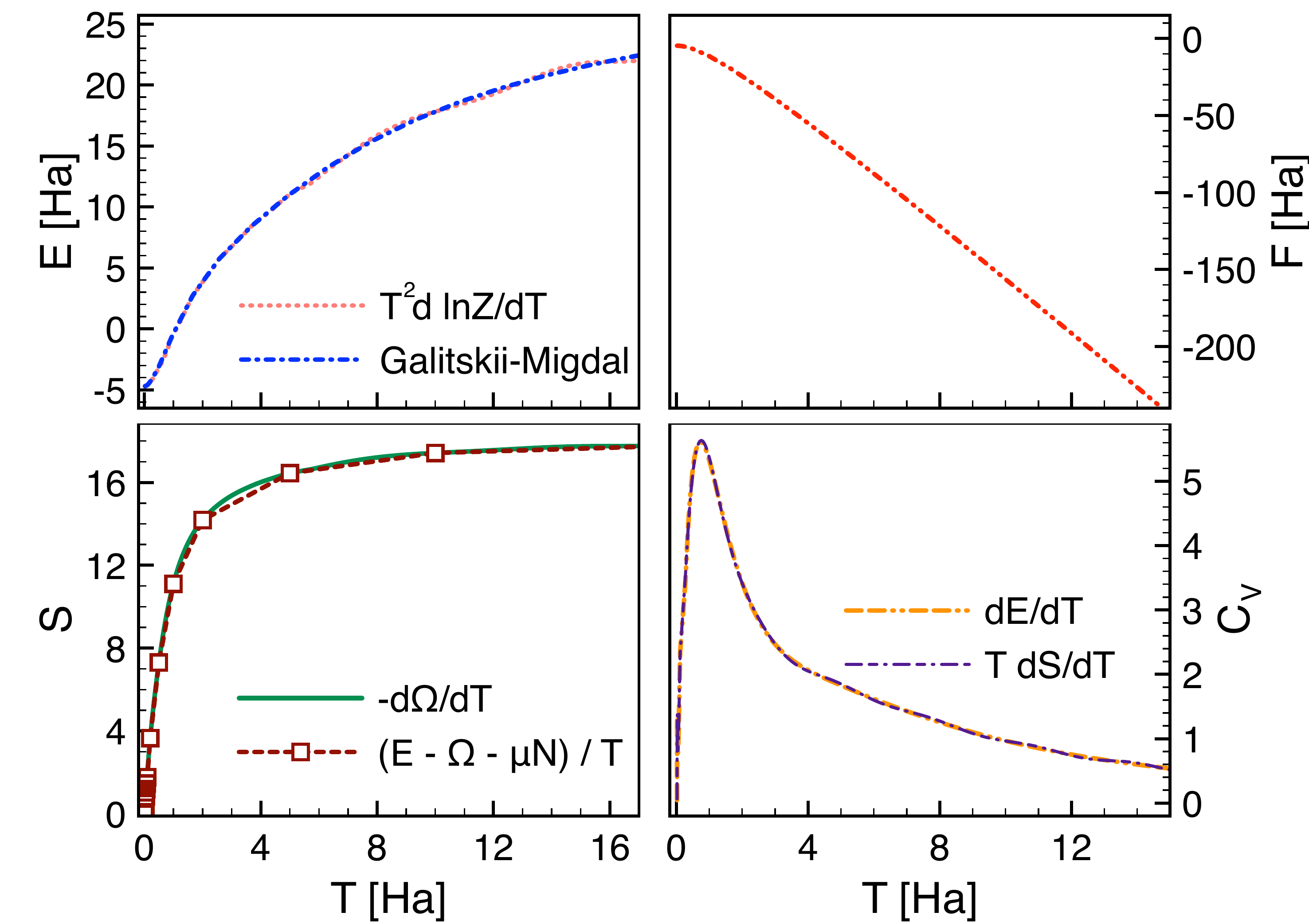}
\caption{Thermodynamics of solid LiH. Clockwise from the top: Internal energy, Helmholtz energy, Specific heat, and entropy as a function of temperature. Data evaluated on a periodic 4$\times$4$\times$4  lattice in the pob-TZVP~\cite{doi:10.1002/jcc.23153} basis.}
\label{Entropy}
\end{figure}

\begin{figure*}[t]
\includegraphics[width=\textwidth]{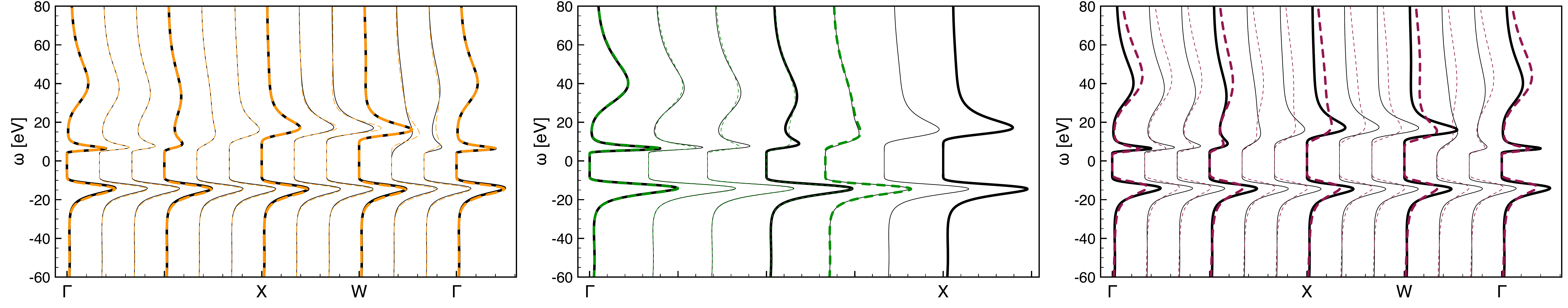}
\caption{Momentum-resolved spectral function for solid neon obtained at $\beta=100$ Ha $^{-1}$. Left panel: comparison between self-consistent GF2 (black lines) and bare second-order perturbation theory (orange lines) on a periodic 4$\times$4$\times$4 lattice. 
Thick lines: directly evaluated k-points. Thin lines: interpolation. Results are listed in the 6-311+G* basis set. Middle panel:
Comparison of the momentum resolved GF2 spectral function along the $\Delta$ direction on a periodic 4$\times$4$\times$4 (solid black) and 3$\times$3$\times$3 (dashed green) lattice. 
Right panel: GF2 on a periodic 4$\times$4$\times$4 lattice, in the basis set 6-311+G* (black) and aug-cc-pVDZ (purple).}
\label{fig:Ne}
\end{figure*}

\section{Results}
We analyze five solids in this paper: Ne, LiF, MgO, LiH, and diamond. The experimental band gaps of these solids are listed in Table~\ref{tab:cmp}, along with literature values obtained by other methods. To confirm the thermodynamic consistency of our implementation, we present the evaluation of thermodynamic properties for solid LiH in Fig.~\ref{Entropy}. Smooth curves have been obtained by Chebyshev interpolation on a temperature grid.

As these insulators have such vastly different band gaps, a different amount of Green's function renormalization is expected to be necessary, allowing us to examine how the iterative nature of GF2 changes the results in comparison to bare perturbation theory and Hartree-Fock. Moreover, since evaluating band gaps using either the QP equation as used in Ref.~\onlinecite{Kresse_solidsII} or analytical continuation \cite{jarrell1996bayesian} may give different results, we compare our values for band gaps (obtained using analytical continuation) to the ones available in the literature, where QP equations were used with the bare perturbation theory.

Our GF2 implementation for periodic systems uses a compact Chebyshev polynomial~\cite{Gull18} representation of Green's functions that converges exponentially (for alternative techniques see~\cite{Kananenka15,Kananenka16}) and relies on the open source ALPS library~\cite{Gaenko17}.
We use periodic density-fitted integrals in Gaussian orbitals, evaluated using the open source pySCF package \cite{McClain17,PYSCF}.
Our calculations result in a set of imaginary time self-energies on discrete $k$-points in the Brillouin zone. In order to obtain smooth spectral functions, we perform a three-dimensional periodic spline interpolation \cite{doi:10.1137/0717021,Fuchs11}. The validity of this procedure is assessed by repeating calculations on a set of grid sizes and obtaining convergence of the interpolated quantities. Throughout this work we show data at an inverse temperature of $\beta=100$Ha$^{-1}$ ($T \sim 3158$K $\sim 0.27$ eV). This temperature is much lower than other energies in the system, in particular much lower than the gaps of the solids discussed here, so that results can be considered to be close to the ground state.

Solid neon has a large experimental band gap of $21.7$ eV \cite{Boursey70}. Consequently, we do not expect strong `screening' or a substantial renormalization of the Green's function.
The left panel of Fig.~\ref{fig:Ne} shows the momentum-resolved spectral function obtained with bare and self-consistent second order techniques along a standard path in the Brillouin
zone (from $\Gamma$ via $X$ and $W$ back to $\Gamma$) for this system, illustrating that self-consistency leads to a negligible change of the Green's function renormalization.
Data is evaluated in the 6-311+G$^*$ basis of Gaussian orbitals \cite{doi:10.1063/1.438955,doi:10.1002/jcc.540040303}. Thick lines denote points in the Brillouin zone that
coincide with our momentum grid. Thin lines denote interpolated values. All spectral functions are plotted as a function of frequency in eV.

In this system, the band gap and the spectral functions are converged with respect to the momentum discretization. This is illustrated in the middle panel of Fig.~\ref{fig:Ne},
where we show results on a 3$\times$3$\times$3 and on a 4$\times$4$\times$4 lattice along the $\Delta$ direction in momentum space. Thick lines denote values on the respective
momentum grids (black for 4$\times$4$\times$4, green for 3$\times$3$\times$3), thin lines are obtained by interpolation. The $X$ point is absent on the smaller grid. 
Data has been obtained in the 6-311+G* basis set.

Our calculations also demonstrate that, in the frequency window shown, the spectral function is relatively insensitive to the choice of the basis set, see right panel of Fig.~\ref{fig:Ne}.
However, it should be stressed that while our results do not show significant differences between aug-cc-pVDZ \cite{doi:10.1063/1.456153} and 6-311+G$^*$ basis sets, they may not be converged
with respect to the basis set size, since both bases are small. Converging our calculations with respect to the basis set size would require a systematic increase of the cardinal number $X$
in the series of aug-cc-pV$X$Z basis sets. This is exceedingly difficult in ordinary solid state calculations since regular Gaussian basis sets such as aug-cc-pV$X$Z become linearly dependent for higher cardinal numbers.

\begin{figure*}[bth]
\includegraphics[width=\textwidth]{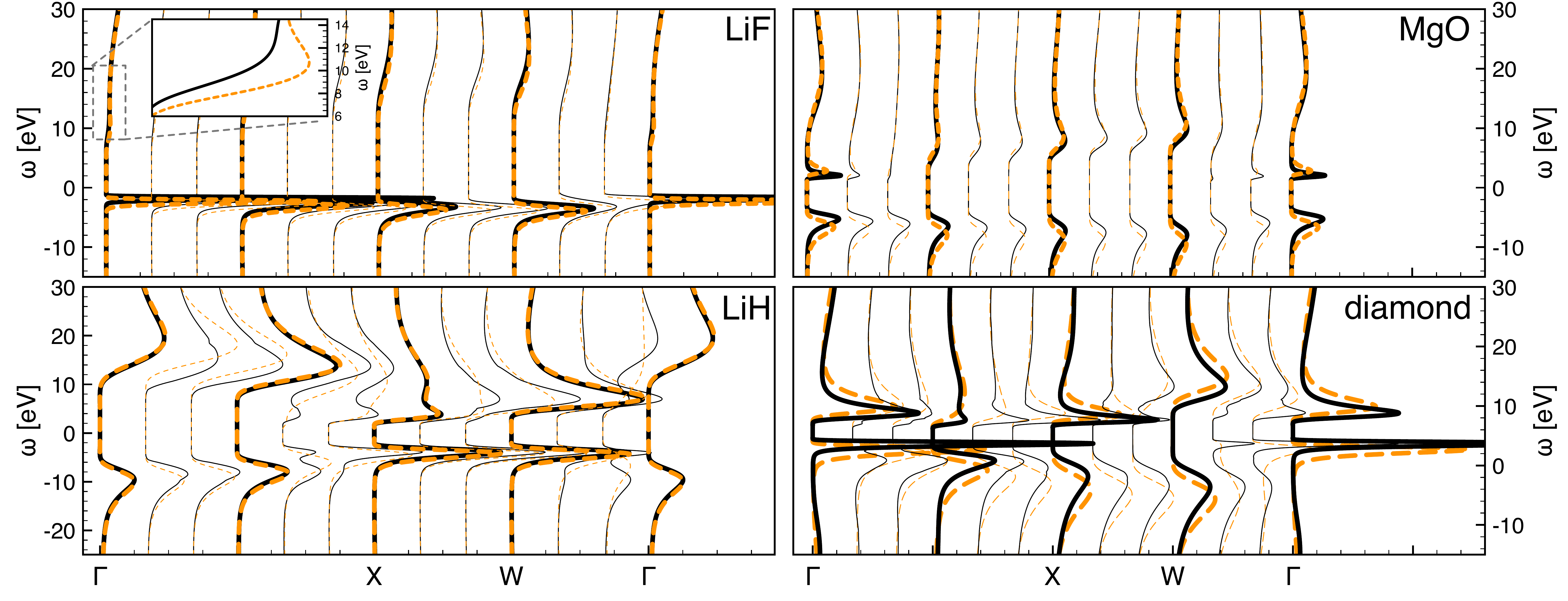}
\caption{Momentum-resolved spectral function of solid LiF (top left), MgO (top right), LiH (bottom left), and diamond (bottom right) obtained from GF2 (black lines) and bare second-order perturbation theory (orange lines)
on a periodic 4$\times$4$\times$4 lattice. Results are listed in the pob-TZVP basis. Inset: zoom to the upper gap edge as indicated in the main panel.}
\label{fig:Bands}
\end{figure*}

Hartree-Fock gaps can be extracted directly from the eigenvalues. The determination of the gap in correlated methods via the spectral function leaves some arbitrariness, as finite temperature Green's functions
are broadened by self-energy, temperature, and analytical continuation artifacts. For GF2, we chose to define the band gap as the peak-to-peak distance of the k-space peaks closest to the Fermi energy from 
above and below. For solid neon, GF2 in both the aug-cc-pVDZ and the 6-311+G$^*$ basis yields a band gap of $20.55$ eV at the $\Gamma$ point. We find that the difference between the GF2 gap ($20.55$ eV) and 
the experimental band gap ($21.7$ eV) is consistent with the previous bare perturbation theory studies\cite{Kresse_solidsII}.

Solid LiF also has a wide experimental band gap of $14.2$ eV \cite{PhysRevB.13.5530}. For the 4$\times$4$\times$4 k-point grid, we observe a band gap of $13.03$ eV in self-consistent GF2,
as extracted from the peak-to-peak distance of the spectral function. The difference of the spectral function between the first iteration and self-consistent GF2 is negligible, see left panel of
Fig.~\ref{fig:Bands}. For k-grid convergence of the unit cell energy and band gap see Tab.~\ref{tab:LiF} and Fig.~\ref{fig:LiF} in appendix~\ref{appx:LiF}.

Both LiH and diamond have moderate band gaps. Therefore, a significant change of the band gap between the first iteration of GF2 and the fully self-consistent result is expected. 
Indeed, both the middle and the right panel of Fig.~\ref{fig:Bands} confirm that  the band gaps obtained in the initial GF2 iteration are much wider than the self-consistent result, and gradually shrink during the  self-consistent iteration progress. Here, to evaluate spectral functions during the first iteration of GF2, we use the analytical continuation of $G(i\omega)=[(i\omega+\mu)1-F_\text{HF}-\Sigma^{(2)}]^{-1}$,
where $\Sigma^{(2)}$ was obtained in the first iteration of the GF2 method, {\it i.e.} using $G_\text{HF}(i\omega)=[(i\omega+\mu)1-F_\text{HF}]^{-1}$ as the propagator in Eq.~\ref{eq:sigma2}.
The Fock matrix $F_\text{HF}$ comes from a preceding HF calculation. Consequently, the first iteration of GF2 lacks two types of renormalization: first, the renormalization coming from the 
self-consistently updated $\Sigma_\infty$ and then consequently updated Fock matrix and, second, the one from the fully self-consistent evaluation of $\Sigma^{(2)}$ which in subsequent
iterations is evaluated with renormalized propagators. 

For diamond in the pob-TZVP basis, we observe an indirect band gap of $4.8$ eV between $\Gamma$ and the point halfway between the $\Gamma$ and $X$ points. The direct band gap at the $\Gamma$-point is about $6.6$ eV.
The positions and values of the direct and indirect band gaps are in good agreement with previous experimental and theoretical results.~\cite{PhysRev.136.A1445,PhysRev.170.683,Clark312,PhysRevLett.16.354}
It is worth emphasizing that the band gap for diamond obtained from self-consistent GF2 is $4.8$ eV, while the MP2 band gap obtained using a QP formalism listed in Ref.~\onlinecite{Kresse_solidsII} is $1.9$ eV.
This underestimation of band gaps smaller than $6$ eV is very noticeable (as listed in Ref.~\onlinecite{Kresse_solidsII}) for bare second order perturbation theory coupled with band gap evaluation based on the QP formalism.
This deficiency seems to be avoided when the fully self-consistent GF2 is employed, as we observe both in the case of LiH and of diamond.

In order to compare results to implementations of MP2, scGW, and QSGW, we also present results for MgO, which has an experimental band gap of $7.8$ eV. 

\begin{figure}[htp]
\includegraphics[width=.95\columnwidth]{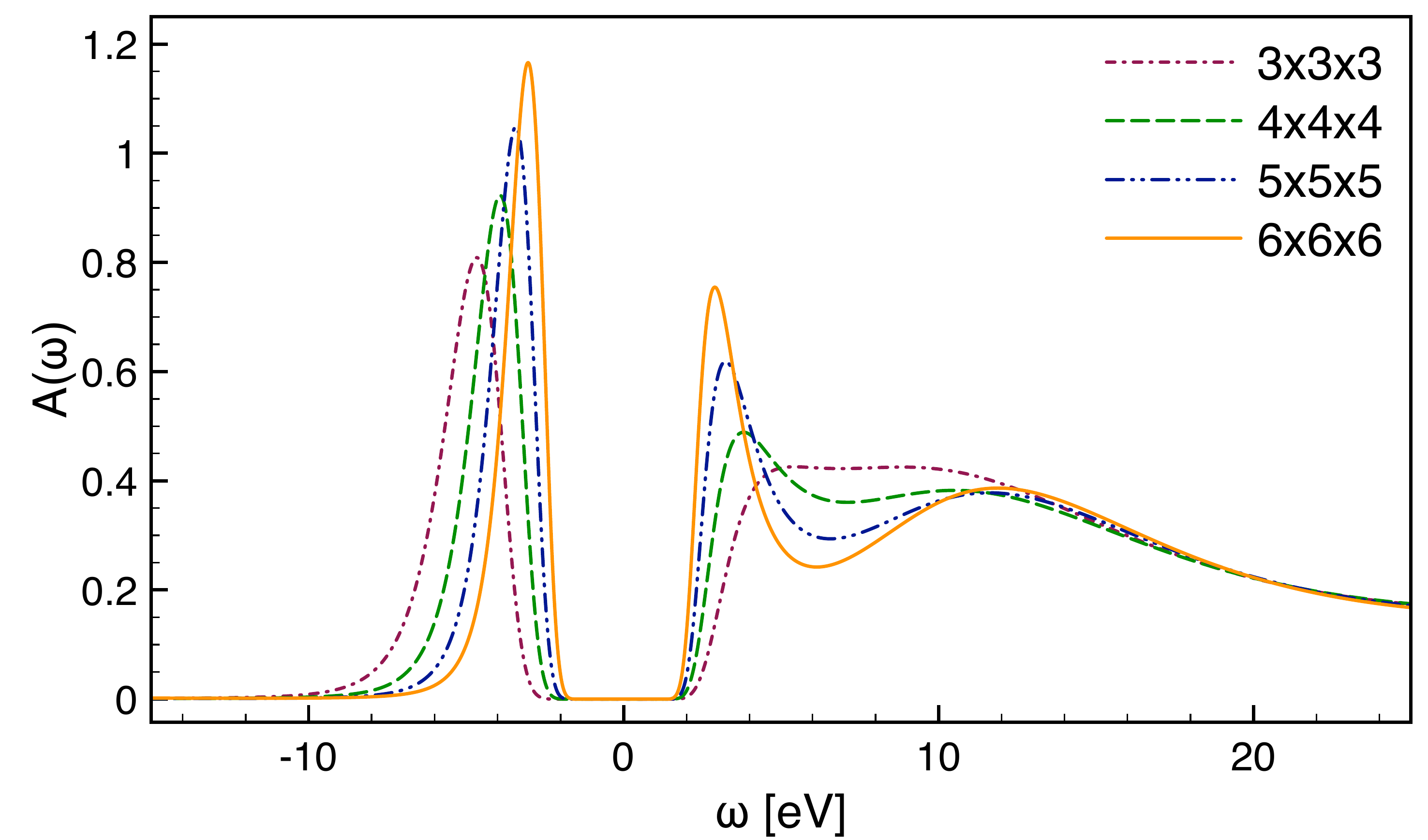}
\caption{Gap size in solid LiH for different k-grids. Results are listed in the basis pob-TZVP.}
\label{fig:Gap_LiH}
\end{figure}

For LiH, we plot the convergence of the band gap and unit cell energy for different k-grids in Tab.~\ref{tab:LiH} and in Fig.~\ref{fig:Gap_LiH}. The data clearly show that while the unit cell energy converges rather quickly, 
the convergence of the band gap is less rapid. While the width of the obtained band gap is bigger than the experimentally observed value, we did not achieve convergence with system size.
An extrapolation of the gap with inverse system size yields a gap value close to $4.0$ eV.

\begin{table}[bth]
\begin{tabular}{ | c || c | c || c | c |}
Size & E (HF) & $e_g$ (HF) & E (GF2)& $e_g$ (GF2)\\ 
 \hline
3$\times$3$\times$3 & -8.0659 & 11.82 & -8.1097 & 10.08\\
4$\times$4$\times$4 & -8.0629 & 11.44 & -8.1079 & 7.73\\
5$\times$5$\times$5 & -8.0618 & 11.31 & -8.1076 & 6.66\\
6$\times$6$\times$6 & -8.0612 & 11.24 & -8.1075 & 5.93\\
\end{tabular}
\caption{\label{tab:LiH}HF and GF2 total energies (E) and band gaps ($e_g$) of LiH for different system sizes in pob-TZVP basis}
\end{table}

\section{Conclusions}
In conclusion, we  illustrated the effect of propagator renormalization on the example of gaps and momentum-resolved spectral functions for Ne, LiH, MgO, LiF, and diamond. In all of these cases, we found reasonable agreement of GF2 with experimental values. Wide-gap insulators were found to be fully converged w.r.t the k-point grid, whereas larger momentum grids are needed to converge the band gap (but not the total energy) of LiH. We showed that thermodynamic consistency is obeyed in our calculations, opening the door for systematic thermodynamic calculations of the electronic system of real materials.

Our study reveals three major results. First, the comparison between our results for bare perturbation theory and self-consistent perturbation theory to experiment shows that renormalized propagator diagrams are responsible for most of the difference between bare results and the experiment. This illustrates the importance of propagator renormalization to obtain the reasonable band gaps for insulators.
Second, the fact that our results from bare second order perturbation theory yield reasonable gap values that differ substantially from the published MP2 values indicates a breakdown of the QP formalism for gap extraction. The approximations inherent to this formalism allow the expression of the results in a convenient real-frequency `band' picture but, in light of our discrepancies, will need to be revisited.
Finally, our results show that controlled self-consistent diagrammatic many-body calculations in standardized Gaussian basis sets are now routinely possible.

\begin{acknowledgments}
S.I. and E.G. acknowledge support by the Simons Foundation via the Simons Collaboration on the Many-Electron Problem. D.Z. and A.R. are supported by NSF-CHE-1453894. E.G and D.Z thank the Simons Foundation for sabbatical support. The Flatiron Institute is a division of the Simons Foundation. S.I. thanks Miguel A. Morales for fruitful discussion. This research used resources of the National Energy Research Scientific Computing Center, a DOE Office of Science User Facility supported by the Office of Science of the U.S. Department of Energy under Contract No. DE-AC02-05CH11231.
\end{acknowledgments}

\appendix
\section{System-size convergence for LiF\label{appx:LiF}}
\begin{figure}[!htp]
\includegraphics[width=.95\columnwidth]{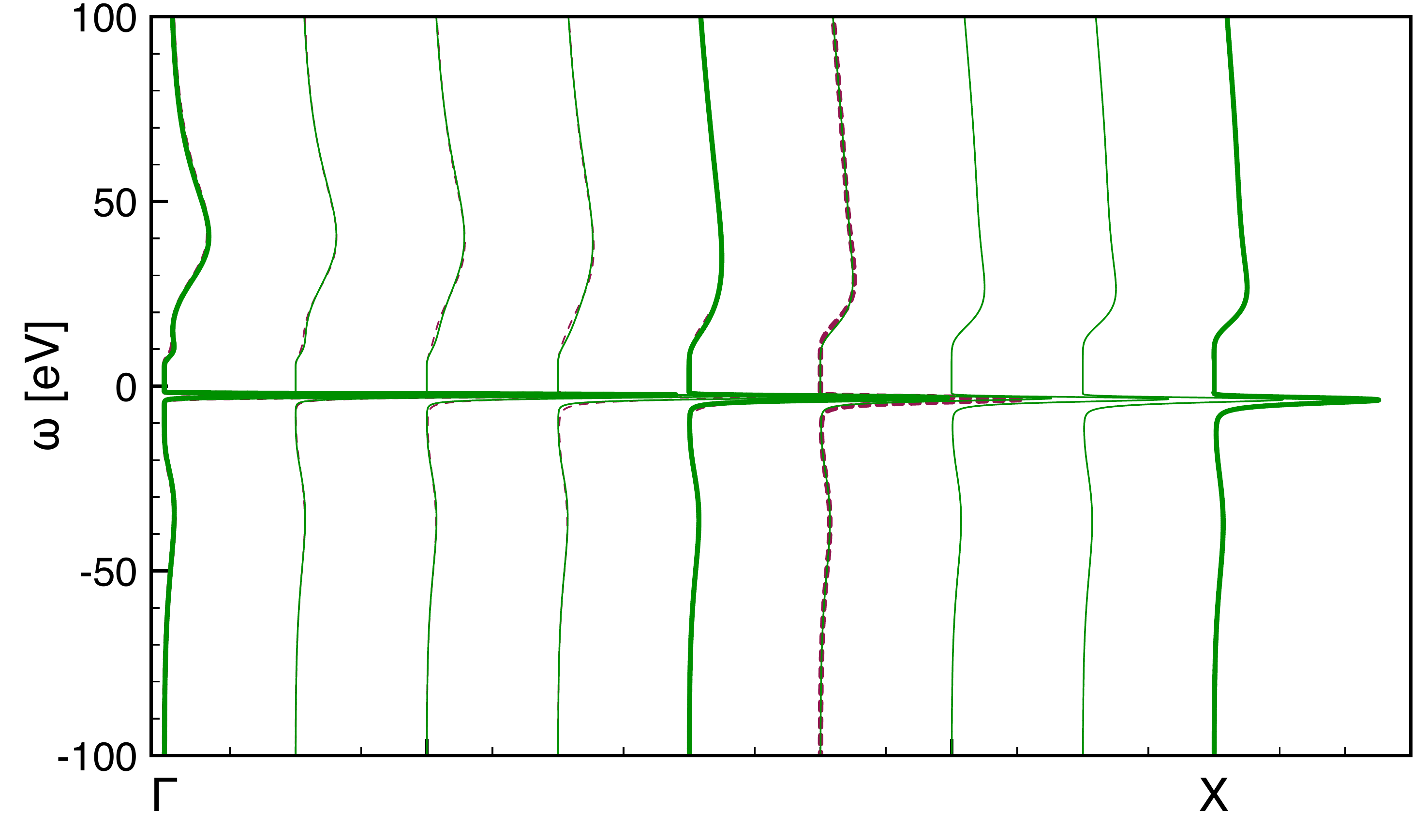}
\caption{Band structure for the LiF obtained from GF2 calculations on 3$\times$3$\times$3 (dashed red) and 4$\times$4$\times$4 (green) periodic lattices in pob-TZVP basis set.}
\label{fig:LiF}
\end{figure}

Tab.~\ref{tab:LiF} shows convergence of the unit cell energy and band gap with momentum grid. Fig.~\ref{fig:LiF} shows the band structure obtained from two momentum grids.

\begin{table}[htp]
\begin{tabular}{ | c || c | c || c | c |}
Size & E (HF) & $e_g$ (HF) & E (GF2)& $e_g$ (GF2)\\ 
 \hline
3$\times$3$\times$3 & -107.092 & 21.95 & -107.325 & 13.36 \\
4$\times$4$\times$4 & -107.089 & 21.84 & -107.321 & 13.03 \\
\end{tabular}
\caption{\label{tab:LiF}HF and GF2 total energies (E) and band gaps ($e_g$) of LiF for systems of size 3$\times$3$\times$3 and 4$\times$4$\times$4 in pob-TZVP basis.
}
\end{table}

\section{Evaluation of thermodynamic properties\label{appx:GP}}

The grand potential is defined in terms of Green's functions, self-energies, and a $\Phi$-functional as \cite{Luttinger60}
\begin{align}
\Omega &= \frac{1}{\beta} \left\{ \Phi[G] - \Tr{\ln\left[-G^{-1}\right]} - \Tr{\Sigma G}  \right\}. \label{appx:GrandPot}
\end{align}
In practice, the evaluation direct evaluation of the second term in this form is complicated by the slow decay of Green's functions as a function of frequency~\cite{Dahlen06}.
We therefore by defining
\begin{subequations}
\begin{align}
G^{-1} &= (i\omega_n  + \mu) S - F - \Sigma_c\left[G\right] \label{appx:bold},\\
G_{HF}^{-1} &= (i\omega_n  + \mu) S - F  \label{appx:HF},\\
G_{0}^{-1} &= (i\omega_n  + \mu) S - H_{0} \label{appx:bare},
\end{align}
\label{appx:GreensFunctions}
\end{subequations}
\noindent
in terms of Matsubara frequencies $i\omega_n$, the chemical potential $\mu$, the overlap matrix $S$, the non-interacting $(V=0)$ Hamiltonian $H_0$ and the Fock matrix $F$
can evaluate the logarithmic term as~\cite{Dahlen06}
\begin{align}
\Tr{\ln\left[-G^{-1}\right]} = \Tr{\ln\left[\Sigma_c-G_{HF}^{-1}\right]}=& \nonumber\\
         \Tr{\ln\left[(-G_{HF}^{-1})(-G_{HF} \Sigma + 1)\right]} =& \nonumber \\
         \Tr{\ln\left[-G_{HF}^{-1}\right]} + \Tr{\ln\left[1 - G_{HF} \Sigma\right]}&. \label{appx:Log}
\end{align}
The $\Phi$-functional is expressed as
\begin{align}
\Phi\left[G\right] = \sum_{n = 1}^{\infty} \Phi^{(n)}\left[G\right]
 = \sum_{n = 1}^{\infty}\frac{1}{2n} \Tr{\Sigma^{(n)}\left[G\right] G}, \label{appx:Phi_n}
\end{align}
where $\Sigma^{(n)}\left[G\right]$ is the total n-th order self-energy part. Within the second-order theory, the self-energy is approximated as  $\Sigma \approx \Sigma^{[2]} = \Sigma^{(1)} + \Sigma^{(2)}$,
and using Eq.~(\ref{appx:Phi_n}) $\Phi \approx \Phi^{[2]} = \Phi^{(1)} + \Phi^{(2)} = \frac{1}{2}\Tr{\Sigma^{(1)}\gamma} + \frac{1}{4}\Tr{\Sigma^{(2)} G}$. Using
Eqs.~(\ref{appx:GreensFunctions}) $\Sigma^{(1)} = F - H_{0}$, such that
\begin{equation}
 \Tr{\Sigma\left[G\right] G}= \Tr{\Sigma^{(1)}\gamma} + \Tr{\Sigma^{(2)} G}. \label{appx:trace}
\end{equation}

Combining Eq.~\ref{appx:Log}-Eq.~\ref{appx:trace} we obtain the second order approximation of the grand potential as
\begin{widetext}
\begin{align}
\Omega^{[2]} &= \frac{1}{\beta} \left\{ \Phi^{[2]}[G] - \Tr{ln\left[-G^{-1}\right]} - \Tr{\Sigma^{[2]} G}  \right\} =\nonumber\\
      &= \frac{1}{\beta} \left\{ \frac{1}{2}\Tr{\Sigma_{\infty}\gamma} + \frac{1}{4}\Tr{\Sigma^{(2)} G} -
         \Tr{ln\left[-G_{HF}^{-1}\right]} - \Tr{ln\left[1 - G_{HF} \Sigma^{(2)}\right]} - \Tr{\Sigma^{(1)}\gamma} - \Tr{\Sigma^{(2)} G} \right\} =\nonumber \\
      &= \frac{1}{\beta} \left\{ -\frac{1}{2}\Tr{\Sigma_{\infty}\gamma} - \frac{3}{4}\Tr{\Sigma^{(2)} G} - \Tr{ln\left[-G_{HF}^{-1}\right]} - \Tr{ln\left[1 - G_{HF} \Sigma^{(2)}\right]} \right\}
\end{align}
\end{widetext}
Standard textbook relations yield the entropy, specific heat, total energy and free energy as thermodynamic derivatives
\begin{subequations}
\begin{align}
    &S = -\frac{\partial \Omega}{\partial T},\\
    &C_V = T \frac{\partial S}{\partial T}, \\
    &E = T^2\frac{\partial ln Z}{\partial T}, \\
    &F = \Omega + \mu N
\end{align}
\end{subequations}
Alternatively, the entropy can be evaluated from the Gibbs-Duhem relation $\Omega = E - TS - \mu N$; the specific heat from its definition $C_V = \frac{\partial E}{\partial T}$; and the energy from the Galitskii-Migdal formula~\cite{PhysRevB.62.4858}.

\bibliographystyle{apsrev4-1}
\bibliography{refs}

\end{document}